\documentclass[slac_one]{revtex4}
\usepackage{graphicx}
\usepackage{fancyhdr}
\usepackage{subfigure}
\newlength{\ogltemp}

\def\etmiss{\settowidth{\ogltemp}{\big\slash}%
\hspace{0.5mm}\big\slash\hspace{-1.4\ogltemp}{E_T}%
}
\def\ptmiss{\settowidth{\ogltemp}{\big\slash}%
\hspace{0.5mm}\big\slash\hspace{-1.4\ogltemp}{P_T}%
}

\begin{document}

\title{{\small{Hadron Collider Physics Symposium (HCP2008),
Galena, Illinois, USA}}\\ 
\vspace{12pt}
Search for SM Higgs Boson Produced in Association with a Z or a W Boson in events with $\etmiss$ and $b$-jets} 

%

\author{A. Apresyan}
\affiliation{Purdue University, West Lafayette, IN 47907, USA}

\begin{abstract}
  We present a search for the Standard Model Higgs boson produced in
  association with a Z or a W boson, using data collected with the CDF
  II detector at the Tevatron accelerator. A scenario where the Z decays
  into neutrinos or charged leptons originating from the W-decay escape
  detection and the Higgs decays into a $b\overline{b}$ pair is
  considered. Therefore the expected signature is large missing
  transverse energy ($\etmiss$), no isolated leptons, and two $b$-jets.
  We present the preliminary results in this search using 1.7$fb^{-1}$
  of data collected by CDF and the work on future improvements to
  increase the sensitivity of the analysis.
\end{abstract}

\maketitle

\thispagestyle{fancy}


\section{INTRODUCTION} 
In the Higgs mechanism of the Standard Model, the fermions and weak
gauge bosons acquire mass via interaction with the Higgs field which is
described by a complex doublet. Three of the four real fields of the
doublet couple to the SU(2) gauge bosons. The observable quantum of the
fourth Higgs field is called the Higgs boson. The existence of this
undiscovered particle is the cornerstone of the Standard
Model\cite{hunter}.

Direct searches performed with the LEP experiments have constrained the
Higgs mass to be larger than 114.4 GeV at 95\% C.L. \cite{lephiggs}. In
$p\overline{p}$ collisions at the Tevatron, the most probable production
mode of the Higgs is by gluon fusion through a virtual top loop. Around
70\% of the Higgs would decay into two $b$-quarks yielding two $b$-jets in
the final state.  Since the QCD $b$-quark production is an irreducible
background, this analysis would have a low sensitivity. The second most
frequent production mode is when a virtual W or Z decays into a W or Z
and a Higgs. In this case, it is possible to trigger on the decay
products of the W/Z boson and significantly reduce the QCD background.

We are analyzing Z-Higgs and W-Higgs associated productions when the Z
decays into two neutrinos, or the W decays leptonically but the charged
lepton escapes the detection. Because the neutrinos will not be detected
in the calorimeter, either, they lead to an unbalanced transverse energy
sum in the transverse plain ($\etmiss$).

\section{DATA SAMPLE AND EVENT SELECTION}
We use data collected through March 2007, which corresponds to 1.7
$fb^{-1}$ integrated luminosity. The events are collected by CDF II
detector with a trigger that selects events with $\etmiss>25$ GeV at
Level 1 at least two Level 2 clusters with $E_{T}>10$ GeV and
$\etmiss>35$ GeV at Level 3 .

In the first step of the analysis, both the Monte Carlo and real data
events are to pass a set of quality cuts to ensure that the possible
beam and detector effects are removed from the data sample making it
compatible with the simulation. The standard CDF jet clustering
algorithm is used \cite{cdfjets} with a jet cone of radius 0.4. Jet
energies are corrected for calorimeter non-uniformity, non-linearity and
energy loss in the un-instrumented regions of calorimeter and energy
coming from different $p \overline{p}$ interactions during the same
bunch crossing. The $\etmiss$ of the event is then corrected with new
jet energies. 

The trigger efficiency is obtained from data and is used to scale the
signal and Monte-Carlo backgrounds to correct for event loss during data
taking. The efficiency of the two-jet requirement is 100\% if the
offline transverse energy of the most energetic jet is above 35 GeV, the
second most energetic jet is above 25 GeV, and at least one of the jets
has $|\eta|<1.0$. The overall efficiency of the online event selection
is then parameterized by the offline corrected $\etmiss$ and applied on
the Monte Carlo samples providing a proper scaling for the simulated
events.

The final requirement imposed on the data and simulation before
comparing them is the $b$-tag requirement.  We use two categories of
SECVTX $b$-tagging algorithm \cite{secvtx}, tight and loose. The main
difference between the loose and tight tagging algorithms is that the
loose tagger has more efficient track selection. The $b$-tagging
efficiency for tight (loose) tagger is $\sim$ 40\% (50\%) and mistag
rate is $\sim$ 2\% (4\%). In this analysis events are split into two exclusive
categories: events with 2 jets tagged by SECVTX loose algorithm or exactly one jet
tagged by SECVTX tight algorithm.

\subsection{BACKGROUND ESTIMATION}
The backgrounds in this data sample have contributions from the
following processes: QCD multi-jet production, top quark pair and single
production, W or Z boson production with jets and diboson production
(WW,WZ,ZZ). We simulate processes which yield real taggable objects,
that is, when a $b$- or a $c$-quark pair is created.  Events with light
flavor jets with a positive tag are considered to be mistags and are
estimated from the data\cite{secvtx}. The remaining background processes
were generated with {\sc pythia}\cite{pythia} Monte Carlo event
generator passed through CDF II detector simulation.

After defining two control regions in the events passing the basic
selection criteria, the Standard Model background is compared to the data. In the
first control region (CR1) all events with identified leptons are
vetoed, and the azimuthal angular separation between the second leading
jet and the $\etmiss$ is less than 0.4. This control region is dominated
by QCD multi-jet events. The second control region (CR2) contains events
with at least one lepton or isolated track and $\varphi(2^{nd}\: jet,
\etmiss)>0.4$. This region is sensitive to Electroweak processes, and is
used to check the overall shapes and normalizations of the Monte Carlo
simulated processes.

\begin{table}[ht]
  \begin{center}
    \begin{tabular}{|l|c|c|c|c|}
      \hline            & \multicolumn{2}{c|}{Control Region 1} & \multicolumn{2}{c|}{Control Region 2} \\\hline
      Process		& 1 Tight tag              & 2 Loose Tags           & 1 Tight tag              & 2 Loose Tags           \\
      \hline\hline
      QCD h.f.          & $24337.1 \pm 111.4 \pm 5445.4 $ & $ 3768.5 \pm 45.8 \pm 688.2 $  & $  50.7 \pm 5.1 \pm 12.6  $  & $  7.0 \pm 2.0 \pm 1.9 $  \\
      Top               & $7.1     \pm   0.4 \pm    0.8 $ & $    2.3 \pm  0.2 \pm   0.4 $  & $ 134.8 \pm 1.6 \pm 16.4  $  & $ 55.9 \pm 1.0 \pm 9.0 $  \\
      Di-boson          & $1.1     \pm   0.2 \pm    0.2 $ & $    0.1 \pm  0.1 \pm   0.1 $  & $  14.7 \pm 0.8 \pm  2.5  $  & $  1.9 \pm 0.2 \pm 0.4 $  \\
      W + h.f.          & $26.2    \pm   2.7 \pm   11.1 $ & $    1.1 \pm  0.5 \pm   0.4 $  & $  80.5 \pm 4.1 \pm 34.9  $  & $  8.2 \pm 1.3 \pm 3.6 $  \\
      Z + h.f.          & $8.7     \pm   1.2 \pm    3.6 $ & $    0.9 \pm  0.4 \pm   0.4 $  & $  17.5 \pm 1.8 \pm  7.9  $  & $  1.3 \pm 0.5 \pm 0.7 $  \\ 
      Mistag            & $6181.0  \pm  63.6 \pm  498.5 $ & $  415.2 \pm 10.0 \pm  71.1 $  & $  86.5 \pm 4.3 \pm  6.3  $  & $  3.7 \pm 1.2 \pm 0.8 $  \\
      \hline \hline                                                                                                                                        
      Expected          & $30561.2 \pm 5469.7           $ & $ 4188.1 \pm 693.4          $  & $ 384.7 \pm 42.6          $ & $ 77.9 \pm 10.3         $  \\ 
      \hline                                                                                                                                                      
      Observed          & $29431~~~~~~~~~~~~~~          $ & $ 4190~~~~~~~~~~~~          $  & $ 373~~~~~~~~~~~          $ & $ 79~~~~~~~~~~~      $     \\  
      \hline
    \end{tabular}
    \caption{Number of expected background and observed events in the control
      regions}\label{tab:CR_events}
  \end{center}
\end{table}

\begin{figure*}[hb]
  \centering
  \includegraphics[width=7cm]{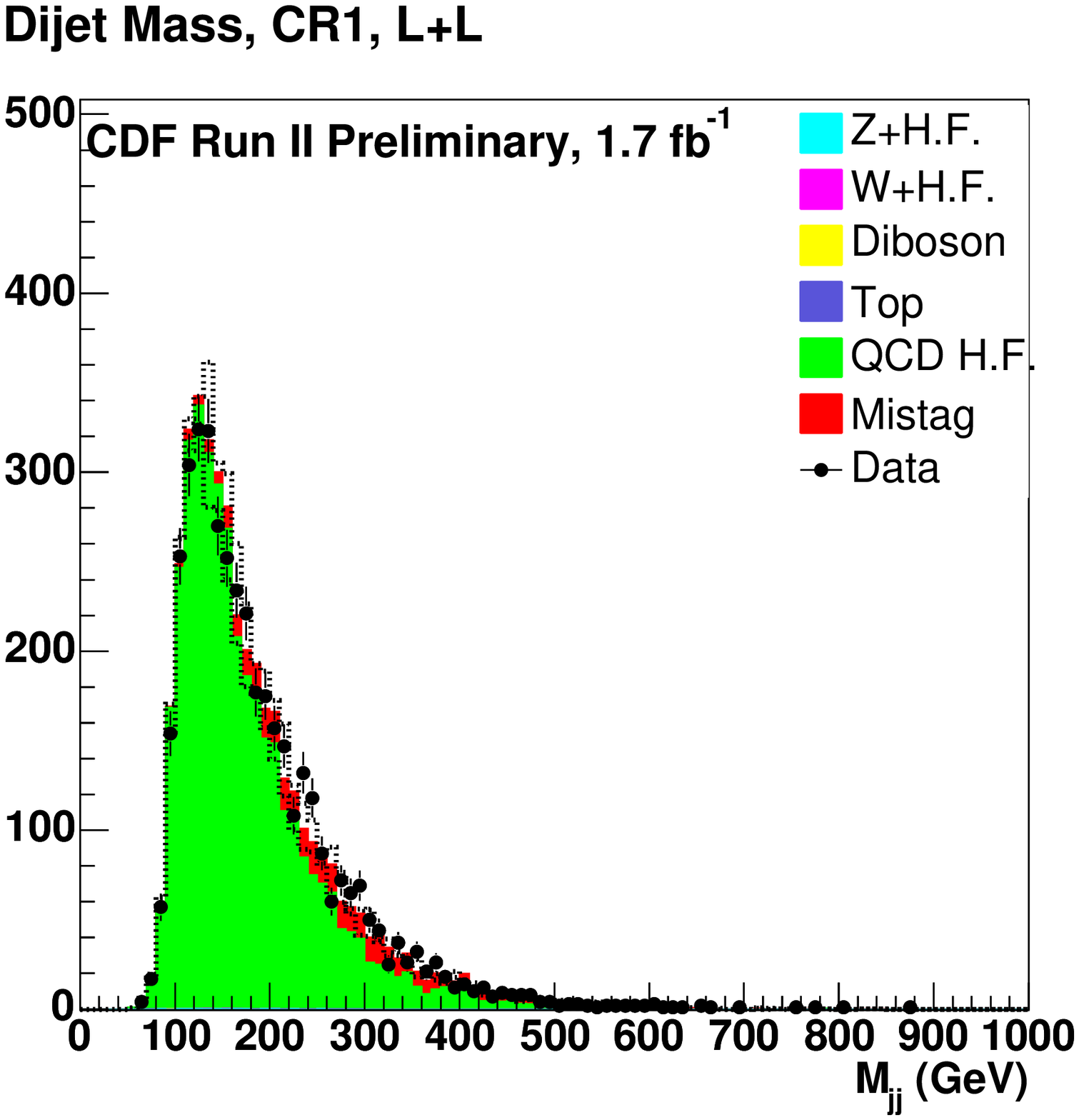}
  \includegraphics[width=7cm]{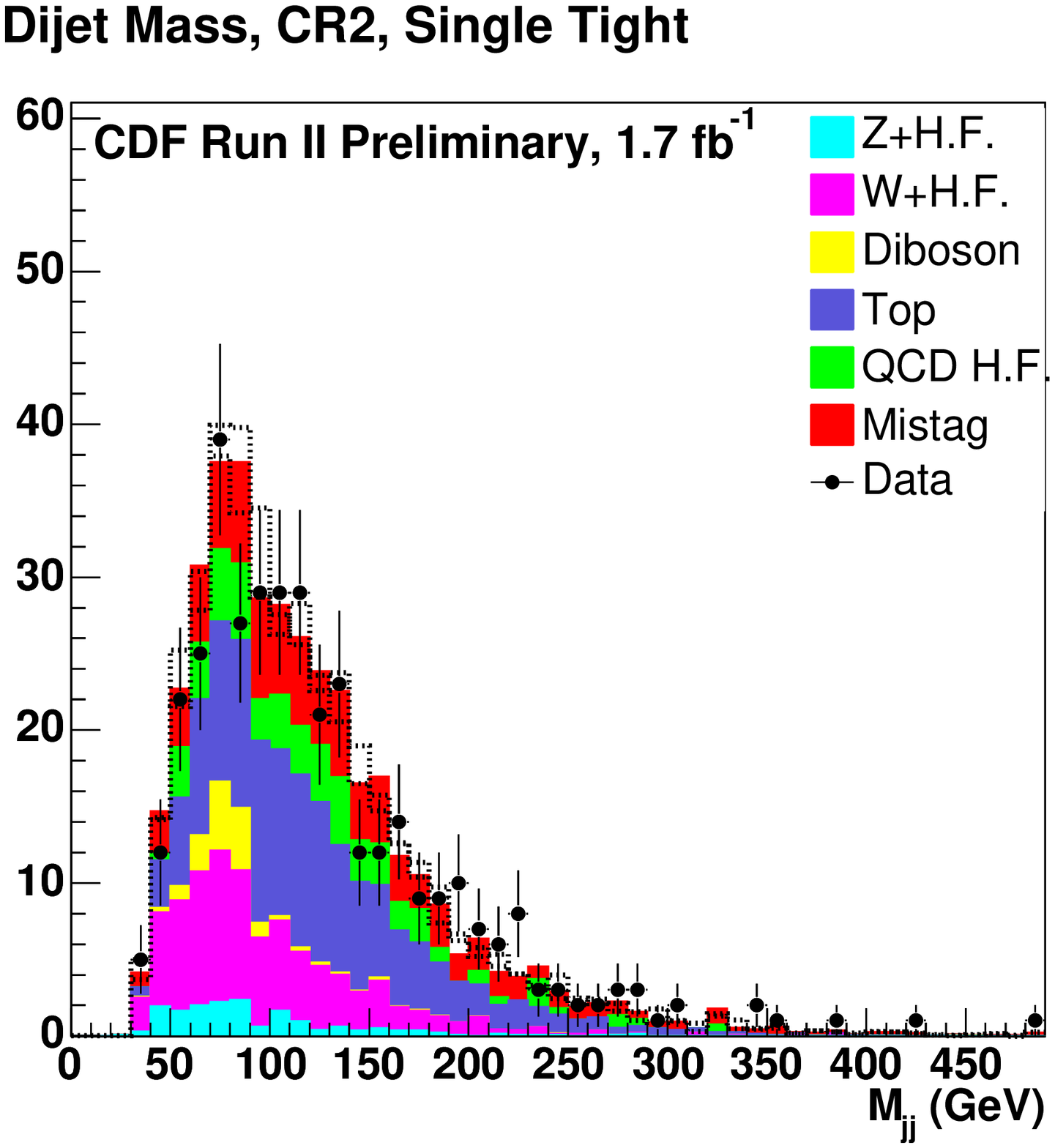}
  \caption{Dijet invariant mass distributions in control regions for:
    (\textit{left}) double-tagged events in CR1, (\textit{right})
    single-tagged events in CR2}
  \label{fig:CR_plots}
\end{figure*}

\begin{figure*}[ht]
  \centering
  \includegraphics[width=7cm]{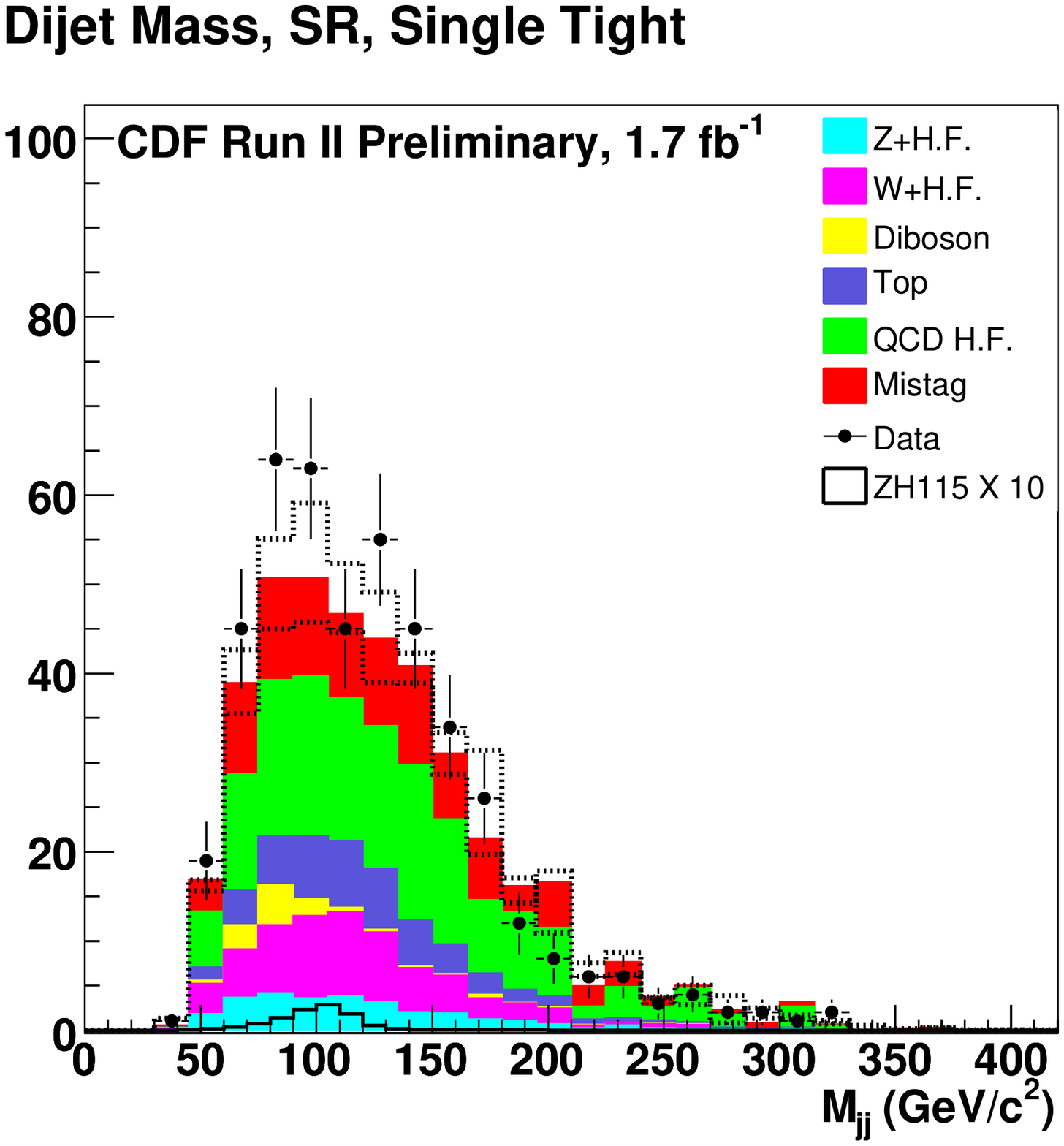}
  \includegraphics[width=7cm]{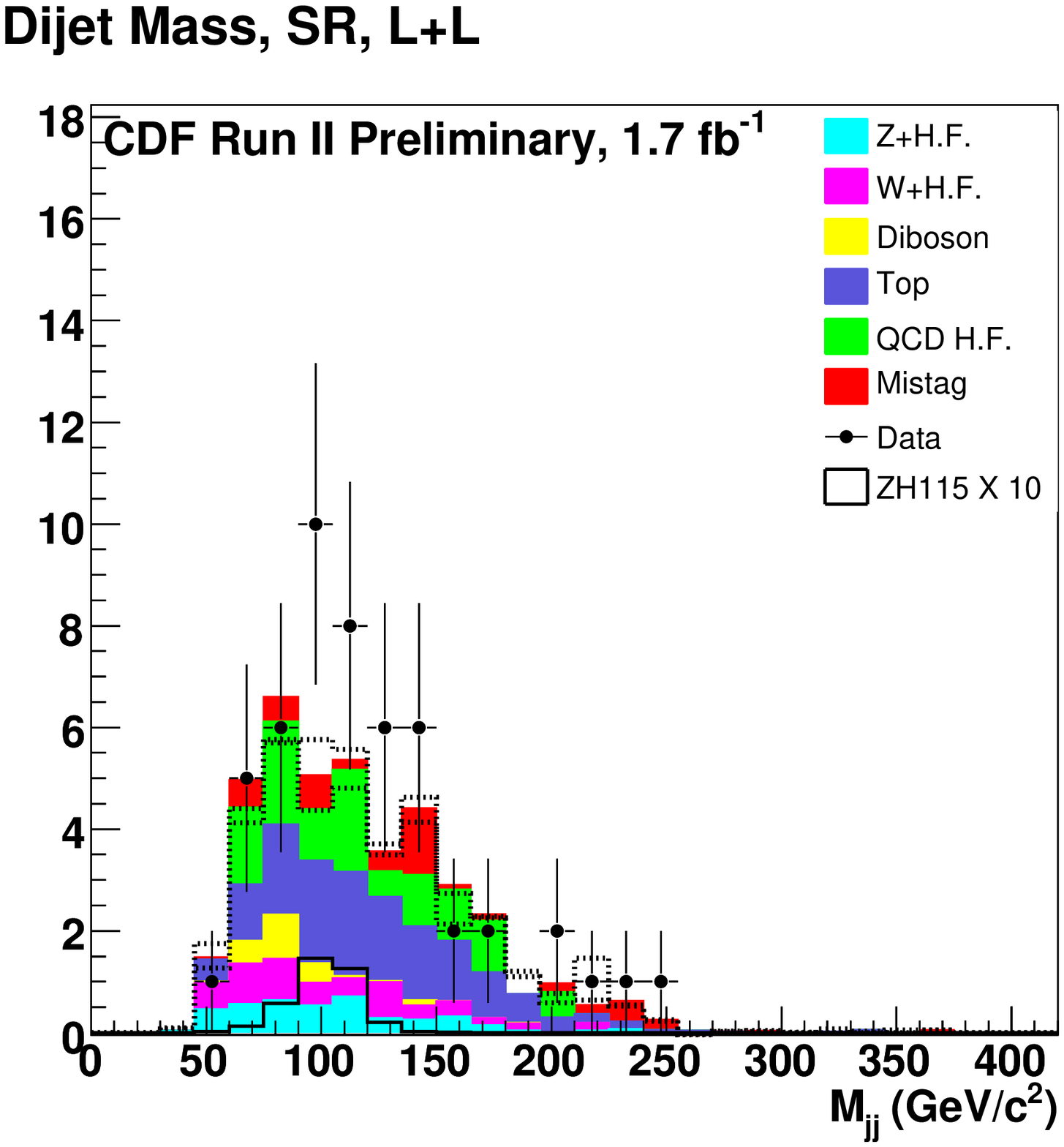}
  \caption{Dijet invariant mass in the Signal Region, single- and
    double-tagged events}
  \label{fig:SR_plots}
\end{figure*}

After achieving a good agreement between the simulation and the data in
the control regions (Table \ref{tab:CR_events}, Figure
\ref{fig:CR_plots}), a set of cuts are selected optimizing the signal
Monte Carlo against the background prediction. The optimization yields
the following selection requirements for signal region (SR):
$\varphi(1^{st} jet, \etmiss)>0.8$, $\varphi(2^{nd} jet, \etmiss)>0.4$,
$\big\slash\hspace{-1.6ex}{H_T}/{H_T}>0.45$\footnote{$H_T$ is the scalar
  sum and $\big\slash\hspace{-1.6ex}{H_T}$ is the magnitude of the
  vectorial sum of the $p_{T}$'s of the two leading jets}, $1^{st} jet
\; E_T > 60$ GeV, $\etmiss>70$ GeV and no isolated leptons in the event.
Table \ref{tab:SR_events} shows the comparison between expected and
observed event yields in the signal region. The dijet invariant mass
distributions in the signal region are shown in Fig.\ref{fig:SR_plots}.
The excess of observed events in double-loose tagged sample has
been extensively studied. No systematic source for disagreement has been
found. The probability of having such an excess as a result of
background only fluctuation was estimated to be $\sim$3\%.

\begin{table}[ht]
  \begin{center}
    \begin{tabular}{|l|c|c|}
      \hline
      Process				    & Single Tag             & Double Tags \\
      \hline\hline
      QCD h.f.          & $ 157.4  \pm  9.0 \pm 49.1$  & $ 10.6 \pm 2.4 \pm 3.9 $\\
      Top               & $  48.2  \pm  1.0 \pm  4.1$  & $ 14.0 \pm 0.5 \pm 2.1 $\\
      Di-boson          & $  11.5  \pm  0.6 \pm  2.4$  & $  1.9 \pm 0.2 \pm 0.4 $\\
      W + h.f.          & $  59.9  \pm  4.1 \pm 26.6$  & $  4.6 \pm 1.1 \pm 2.0 $\\
      Z + h.f.          & $  28.3  \pm  1.9 \pm 12.5$  & $  4.1 \pm 0.7 \pm 1.9 $\\ 
      Mistag            & $  98.2  \pm  7.3 \pm 12.8$  & $  4.7 \pm 1.0 \pm 1.1 $\\\hline \hline
      Expected          & $ 403.5  \pm 60.1         $  & $ 39.9 \pm 6.1         $\\ 
      Observed          & $443~~~~~~~~~~~$                        & $ 51~~~~~~~~~                   $\\  \hline
      $(VH$ $m_H=115\,$GeV/c$^2$) & $(1.9)  $           & $  (1.2)                 $\\\hline
    \end{tabular}
    \caption{Number of expected background and signal events in the
      signal region after applying the final
      cuts.}
    \label{tab:SR_events}
  \end{center}
\end{table}

\subsection{CROSS-SECTION LIMITS}
Since there is no significant excess in the data compared to the
predicted backgrounds, we set 95\% C.L. upper limits on Higgs boson
production cross-section times the branching fraction. The systematic
uncertainties are classified as correlated and uncorrelated errors
considering the relations between the signal and the background
processes. The correlated errors are taken into account separately for
each processes in the limit calculation. The uncorrelated systematic
uncertainties are: statistical error in negative tag estimate,
negative-positive tag rate asymmetry factor, QCD multi-jet Monte Carlo
normalization (14\% in single tagged, 6.3\% in double tagged sample), MC
statistical fluctuations. The correlated systematics are: luminosity
(6.0\%), $b$-tagging efficiency scale factor between data and Monte Carlo
(4.3\% for single and 10.2\% for double tags), trigger efficiency (3\%),
lepton veto efficiency (2\%), PDF uncertainty (2\%) and Jet Energy
Scale. ISR/FSR systematic uncertainties (between 1\% and 5\%) are
applied on the signal.

Considering the systematic uncertainties listed above, we computed the
expected limit for the Higgs cross-section when the Higgs is produced
with a Z/W boson and decays to two $b$-quarks where Z decays to neutrinos
and W to leptons. We use Bayesian method for deriving the
limits\cite{csmcode}. Table \ref{tab:limits} shows the final result.
All the cross-sections are ratios with respect to the Standard Model
cross-section.

\begin{table}
\begin{center}
\begin{tabular}{|l|r|r|r|r|r|r|}
\hline
Higgs mass    & \multicolumn{2}{c|}{VH limit, 1 Tight Tag} 
	            & \multicolumn{2}{c|}{VH limit, 2 Loose Tags}
	            & \multicolumn{2}{c|}{VH limit, Combined} \\
(GeV) & Predicted & Observed & Predicted & Observed & Predicted & Observed \\ \hline \hline

110   & $19.7^{+9.7}_{-6.0}$    & $ 36.6$  & $10.4^{+4.4}_{-2.9}$   & $18.7$  & $9.3^{+4.4}_{-2.9} $    & $18.5$     \\ \hline 
115   & $22.7^{+9.5}_{-7.2}$    & $ 37.2 $ & $11.1^{+4.4}_{-3.3}$   & $20.8$  & $9.7^{+5.0}_{-2.8} $    & $19.7$     \\ \hline 
120   & $27.5^{+11.4}_{-7.7}$   & $ 40.8$  & $13.0^{+6.5}_{-3.9}$   & $25.2$  & $11.5^{+5.5}_{-3.7} $   & $22.6$     \\ \hline 
125   & $31.2^{+14.8}_{-9.3}$   & $ 46.6$  & $15.9^{+6.6}_{-4.8}$   & $30.1$  & $13.4^{+6.1}_{-4.1} $   & $26.6$     \\ \hline 
130   & $40.6^{+16.7}_{-12.6}$  & $ 58.7$  & $19.5^{+10.6}_{-5.5}$  & $39.3$  & $16.6^{+7.3}_{-5.3} $   & $33.4$     \\ \hline 
135   & $52.0^{+22.4}_{-16.7}$  & $ 74.6$  & $24.7^{+10.7}_{-7.5}$  & $48.3$  & $21.0^{+9.7}_{-6.3} $   & $43.0$     \\ \hline 
140   & $71.6^{+31.5}_{-23.7}$  & $ 110.0$ & $35.3^{+17.5}_{-10.9}$ & $64.3$  & $31.5^{+16.4}_{-7.2} $  & $61.5$     \\ \hline 
150   & $172.3^{+71.7}_{-61.4}$ & $ 238.6$ & $77.1^{+37.1}_{-22.4}$ & $133.6$ & $72.1^{+30.9}_{-23.4} $ & $127.0$    \\ \hline 

\end{tabular}
\caption{The predicted and observed cross-section limits of the ZH/WH
  processes combined when $H\rightarrow b\bar{b}$ divided by the SM
  cross-section}\label{tab:limits}
\end{center}
\end{table}

\section{FUTURE PROGRESS}
Improvements of the analysis technique are being developed to further
increase the sensitivity of the analysis. Below we summarize the
progress in the main directions that are pursued.

\begin{figure*}
  \centering \subfigure[Dijet invariant mass, single-tagged events,
  CR2]{\label{fig:CR_imp_plots_mjjj}\includegraphics[width=5cm]{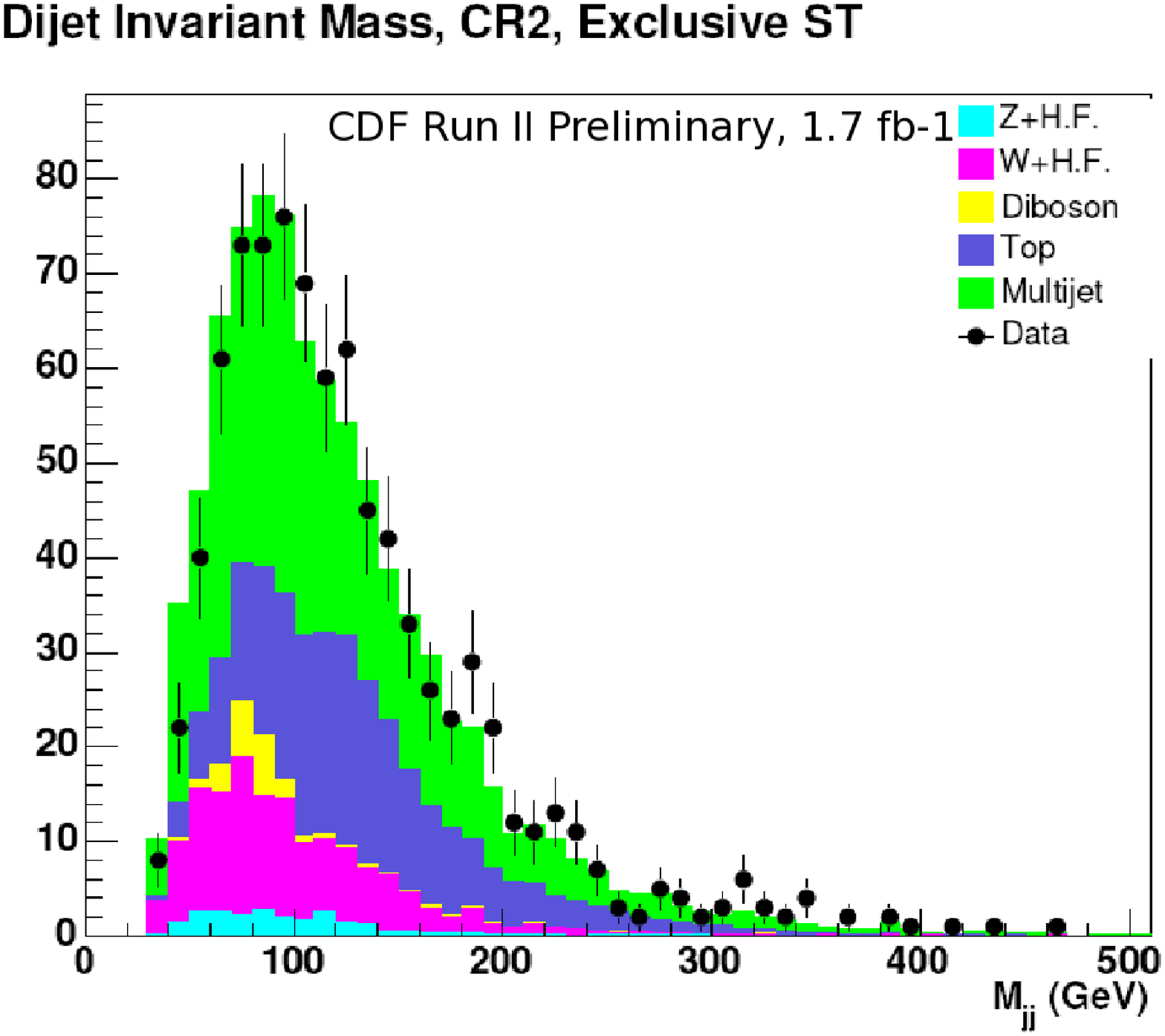}}
  \subfigure[Number of jets, single-tagged events,
  CR2]{\label{fig:CR_imp_plots_njet}\includegraphics[width=5cm]{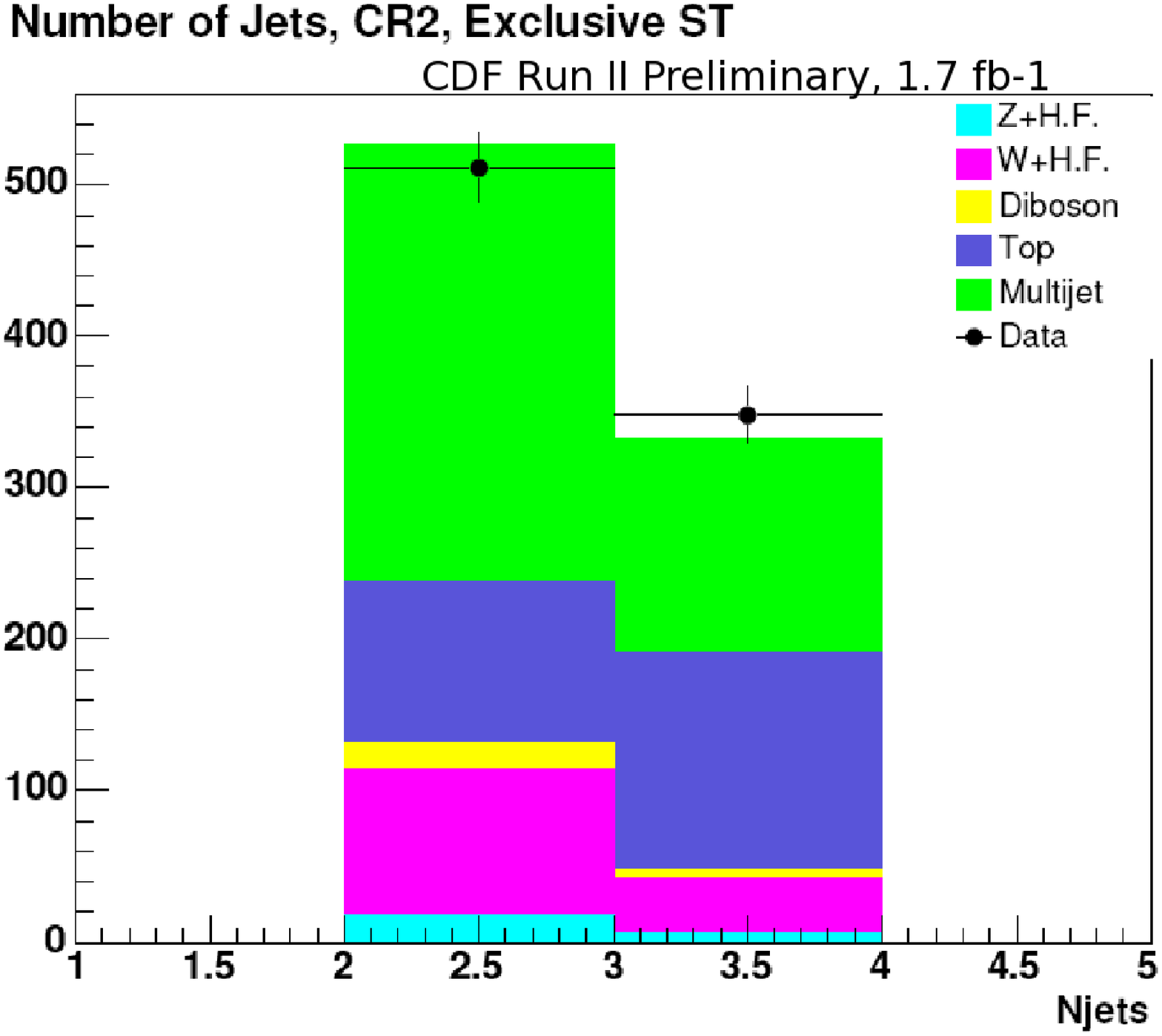}}
  \subfigure[$\Delta \phi (\etmiss,\ptmiss)$, double-tagged events,
  CR1]{\label{fig:CR_imp_plots_mptmet}\includegraphics[width=5cm]{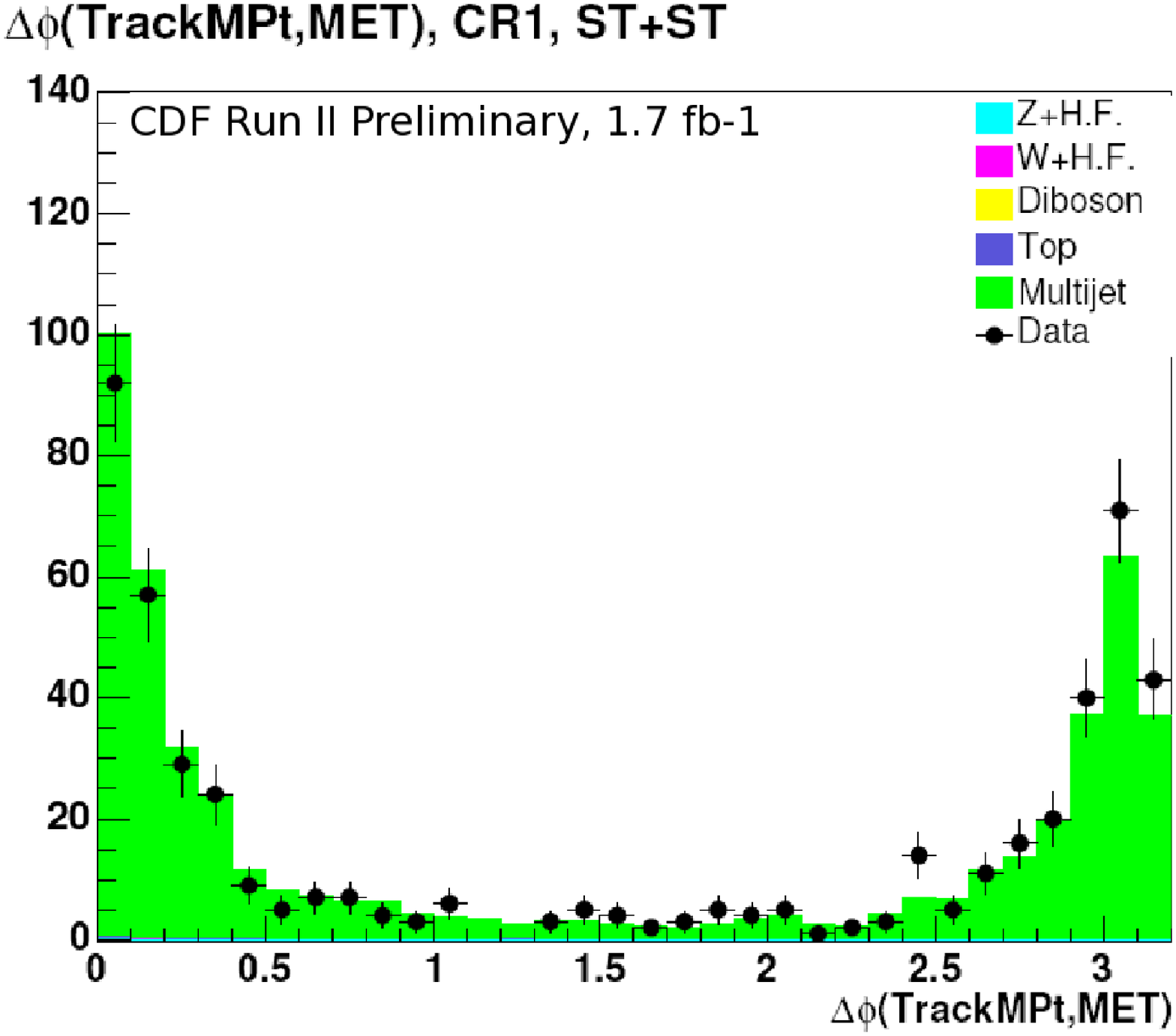}}
  \caption{Kinematic distributions in control regions}
  \label{fig:CR_imp_plots}
\end{figure*}

One of the main challenges in this search channel is the modeling of the
large QCD multi-jet background. A model to estimate this background
directly from data has been developed. In order to estimate the multi-jet
background in the single-tagged sample we measure the probability to tag
one jet from the ``pretag''\footnote{\textit{i.e.} the events passing the basic
selection criteria before the tagging requirement} sample. Similarly, to estimate the multi-jet
background in the double-tagged sample we measure the probability to
tag a jet in a sample that already has one jet tagged. This method
allows us to estimate the shapes and normalizations for multi-jet
production in single-tagged and double-tagged categories,
Fig.\ref{fig:CR_imp_plots_mjjj}.

In order to increase the acceptance to the Higgs signal we accept events
with three jets, for the first time at CDF in this channel. The main
motivation is to accept events where one of the $b$ quarks coming from the
Higgs radiates a gluon. In addition to that, we also accept WH events
where the charged lepton coming from the W is reconstructed as a jet.
The latter case happens when the W decays to an electron and it fails
the CDF electron identification algorithm, but is reconstructed as a
jet; or when the W decays to $\tau \nu$ and $\tau \rightarrow$hadrons.
As can be seen from Fig.\ref{fig:CR_imp_plots_njet} our background
prediction for jet multiplicity agrees very well with the observed data.

In order to effectively reduce the large QCD background, we need to get
a good estimate of the event true missing energy. We do that by
calculating the $\ptmiss$, which is defined as the negative vectorial sum of
charged particle $p_T$'s. For events with true $\etmiss$ the $\ptmiss$
is highly correlated and parallel with calorimeter $\etmiss$, while for
QCD events with mismeasured jets it is not. As shown in
Fig.\ref{fig:CR_imp_plots_mptmet} the $\Delta \phi (\etmiss,\ptmiss)$
can serve as an excellent kinematic variable to discern real from fake
$\etmiss$.

For searches of $H\rightarrow b\overline b$ decays it is crucial to
precisely meausure the energies of the jets coming from the $b$-quarks.
An algorithm similar to the one used by H1 collaboration \cite{H1} has
been implemented, which combines tracks and calorimeter towers to
improve the reconstruction of the jets energies. Initial studies show
that an improvement of $\sim10\%$ in dijet mass resolution can be
achieved. To further increase the final discriminant sensitivity, we are
developing an artificial neural network to maximally separate Higgs
boson signal from backgrounds.

In summary, we have performed a direct search for the Standard Model
Higgs boson decaying into $b$-jet pairs in 1.7 $fb^{-1}$ data
accumulated in Run II of the CDF detector. We do not observe any
significant excess over the background predicted by the Standard Model,
thus we set a 95 \% C.L.  upper limit for the Higgs boson at various
masses. Multiple improvements in the analysis technique are being
developed and will be incorporated in the next iterations of the
analysis.

\end{document}